\documentclass[12pt]{article}
\usepackage{chapterbib}
\usepackage{scicite}
\usepackage{times}
\usepackage{graphicx}
\newcommand{\fs}{\mbox{$.\!\!^{\mathrm s}$}}
\newcommand{\fdg}{\mbox{$.\!\!^\circ$}}
\newcommand{\farcs}{\mbox{$.\!\!^{\prime\prime}$}}
\newcommand{\degrees}{\ifmmode^{\circ}\else$^{\circ}$\fi}
\newcommand{\amin}{\ifmmode^{\prime}\else$^{\prime}$\fi}
\newcommand{\asec}{\ifmmode^{\prime\prime}\else$^{\prime\prime}$\fi}
\newcommand{\psr}{PSR~J1903$+$0327}
\newcommand{\psrnum}{J1903$+$0327}
\newcommand{\ga}{\ifmmode\stackrel{>}{_{\sim}}\else$\stackrel{>}{_{\sim}}$\fi}
\newcommand{\la}{\ifmmode\stackrel{<}{_{\sim}}\else$\stackrel{<}{_{\sim}}$\fi}

\newenvironment{sciabstract}{
\begin{quote} \bf}
{\end{quote}}

\newcounter{lastnote}
\newenvironment{scilastnote}{
\setcounter{lastnote}{\value{enumiv}}
\addtocounter{lastnote}{+1}
\begin{list}
{\arabic{lastnote}.}
{\setlength{\leftmargin}{.22in}}
{\setlength{\labelsep}{.5em}}}
{\end{list}}

\title{\large{An Eccentric Binary Millisecond Pulsar in the Galactic Plane}}

\author{
  \normalsize{David J. Champion$^{1,2\ast}$, Scott M. Ransom$^{3}$, Patrick Lazarus$^{1}$, Fernando Camilo$^{4}$, Cees Bassa$^{1}$,}\\
  \normalsize{Victoria M. Kaspi$^{1}$, David J. Nice$^{5}$, Paulo C. C. Freire$^{6}$, Ingrid H. Stairs$^{7}$, Joeri van Leeuwen$^{8}$,}\\
  \normalsize{Ben W. Stappers$^{9}$, James M. Cordes$^{10}$, Jason W. T. Hessels$^{11}$, Duncan R. Lorimer$^{12}$,}\\
  \normalsize{ Zaven Arzoumanian$^{13}$, Don C. Backer$^{8}$, N. D. Ramesh Bhat$^{14}$, Shami Chatterjee$^{15}$,}\\
  \normalsize{Isma\"el Cognard$^{16}$, Julia S. Deneva$^{10}$, Claude-Andr{\'e} Faucher-Gigu{\`e}re$^{17}$, Bryan M. Gaensler$^{15}$,}\\
  \normalsize{ JinLin Han$^{18}$, Fredrick A. Jenet$^{19}$, Laura Kasian$^{7}$, Vlad I. Kondratiev$^{12}$, Michael Kramer$^{9}$,}\\
  \normalsize{Joseph Lazio$^{20}$, Maura A. McLaughlin$^{12}$, Arun Venkataraman$^{6}$ \& Wouter Vlemmings$^{21}$}\\
  \scriptsize{$^{1}$Dept.~of Physics, McGill Univ., Montreal, QC H3A 2T8, Canada}\\
  \scriptsize{$^{2}$ATNF-CSIRO, PO Box 76, Epping NSW 1710, Australia}\\
  \scriptsize{$^{3}$NRAO, 520 Edgemont Rd., Charlottesville, VA 22903, USA}\\
  \scriptsize{$^{4}$Columbia Astrophysics Laboratory, Columbia Univ., 550 West 120th St., New York, NY 10027, USA}\\
  \scriptsize{$^{5}$Physics Dept., Bryn Mawr College, Bryn Mawr, PA 19010, USA}\\
  \scriptsize{$^{6}$NAIC, Arecibo Observatory, HC03 Box 53995, PR 00612, USA}\\
  \scriptsize{$^{7}$Dept.~of Physics and Astronomy, Univ.~of British Columbia, 6224 Agricultural Rd., Vancouver, BC V6T 1Z1, Canada}\\
  \scriptsize{$^{8}$Astronomy Dept., 441 Campbell Hall, Univ.~of California at Berkeley, Berkeley, CA 94720, USA}\\
  \scriptsize{$^{9}$Jodrell Bank Observatory, Manchester Univ., Macclesfield, Cheshire SK11 9DL, UK}\\
  \scriptsize{$^{10}$Astronomy Dept., Cornell Univ., Ithaca, NY 14853, USA}\\
  \scriptsize{$^{11}$Astronomical Institute ``Anton Pannekoek,'' Univ. of Amsterdam, Kruislaan 403, 1098 SJ Amsterdam, The Netherlands}\\
  \scriptsize{$^{12}$Dept.~of Physics, West Virginia Univ., Morgantown, WV 26506, USA}\\
  \scriptsize{$^{13}$CRESST and X-ray Astrophysics Laboratory, NASA-GSFC, Code 662, Greenbelt, MD 20771, USA}\\
  \scriptsize{$^{14}$Swinburne Univ.~of Technology, PO Box 218, Hawthorn, Victoria 3122, Australia}\\
  \scriptsize{$^{15}$School of Physics, The Univ.~of Sydney, NSW 2006 Australia}\\
  \scriptsize{$^{16}$LPCE / CNRS, UMR 6115 3A, Av de la Recherche Scientifique, F-45071 Orleans Cedex 2, France}\\
  \scriptsize{$^{17}$Harvard-Smithsonian Center for Astrophysics, 60 Garden Street, MS-10, Cambridge, MA 02138, USA}\\
  \scriptsize{$^{18}$National Astronomical Observatories, CAS, Jia-20 DaTun Rd., Chaoyang Dist., Beijing 100012, China}\\
  \scriptsize{$^{19}$Center for Gravitational Wave Astronomy, Univ. of Texas at Brownsville, TX 78520, USA}\\
  \scriptsize{$^{20}$Naval Research Laboratory, 4555 Overlook Ave. SW, Washington, DC 20375, USA}\\
  \scriptsize{$^{21}$Argelander-Institut f\"{u}r Astronomie, University of Bonn, Auf dem H\"{u}gel 71, 53121 Bonn, Germany}\\
  \scriptsize{$^\ast$To whom correspondence should be addressed; E-mail: David.Champion@atnf.csiro.au} 
}

\date{}

\begin{document} 
\baselineskip24pt
\maketitle 

\begin{sciabstract}
  Binary pulsar systems are superb probes of stellar and binary evolution and the physics of extreme environments.  In a survey with the Arecibo telescope, we have found \psr, a radio pulsar with a rotational period of 2.15~ms in a highly eccentric ($e = 0.44$) 95-day orbit around a solar mass companion. Infrared observations identify a possible main-sequence companion star.  Conventional binary stellar evolution models predict neither large orbital eccentricities nor main-sequence companions around millisecond pulsars.  Alternative formation scenarios involve recycling a neutron star in a globular cluster then ejecting it into the Galactic disk or membership in a hierarchical triple system. A relativistic analysis of timing observations of the pulsar finds its mass to be 1.74$\pm$0.04\,M$_\odot$, an unusually high value.
\end{sciabstract}

The population of binary and millisecond pulsars in the disk of our Galaxy is thought to have two main formation mechanisms \cite{Sta04}. Most pulsars with spin periods of tens of milliseconds have neutron star (NS) companions in orbits of high eccentricity caused by the near disruption of the system by a supernova explosion.  In contrast, pulsars with spin periods less than about 10\,ms (i.e.~``millisecond pulsars'' or MSPs) have white dwarf (WD) companions in orbits made highly circular (orbital eccentricities $e < 0.001$) by tidal effects during the recycling process.  The combination of rapid spin rates and circular orbits is considered vital evidence that MSPs achieve their short periods via accretion of mass and angular momentum from binary companion stars \cite{acrs82}.  Here we report the discovery of an unprecedented MSP that requires a different formation mechanism and whose potentially large mass may play an important role in constraining the equation of state of matter at supra-nuclear density \cite{lp07}.

\section*{Discovery and Follow-up Observations}

We are conducting a pulsar survey of the Galactic plane using the Arecibo L-band Feed Array (ALFA) receiver on the 305-m Arecibo radio telescope in Puerto Rico \cite{cfl+06}.  The large collecting area of Arecibo, the rapid sampling rate (every 64\,$\mu$s), and the high spectral resolution (256 channels over 100\,MHz, which minimizes the dispersive smearing due to free electrons along the line of sight) provides sensitivity to MSPs over a much larger volume of the Galactic disk than any previous pulsar survey.

The 2.15-ms pulsar J1903$+$0327 was discovered using a search pipeline based on the PRESTO suite of pulsar analysis software\cite{rem02,presto}. It was detected as a highly significant signal with a large dispersion measure (DM) of 297\,pc\,cm$^{-3}$ in data taken in October 2005. Follow-up timing observations using Arecibo, the Green Bank Telescope, and the Westerbork Synthesis Radio Telescope revealed the binary orbit of the pulsar to be highly eccentric (Fig.~1). The Keplerian orbital parameters give a minimum companion mass of 0.85$-$1.07\,M$_\odot$ (for pulsar masses of 1.3$-$1.9\,M$_\odot$).  Additional constraints come from the extensive timing of the pulsar, including a general relativistic interpretation of the measurement of the advance of periastron, $\dot \omega$, and a detection of the Shapiro delay.  As described in the supplementary online material, and in the caption to Table 1, the best model fit to the pulse times of arrival indicates that the companion has a mass of $1.051(15)$\,M$_\odot$ and that the pulsar has a mass of $1.74(4)$\,M$_\odot$. It should be noted that these masses are based on $\sim$1.5 years of timing data which cannot yet include parameters such as proper motion which can affect the mass measurements. While the pulsar mass is  significantly more massive than the 1.25$-$1.45\,M$_\odot$ seen in most double neutron star (DNS) systems \cite{Sta04} it is comparable to the inferred masses of several recently detected pulsars in eccentric binaries in globular clusters \cite{rhs+05,frb+08,fwvh08}, at least one other Galactic MSP \cite{vbv08}, and the X-ray pulsar Vela~X-1 \cite{qna+03}.  If the large pulsar mass is confirmed in future observations, it will constrain the equation of state of matter at supra-nuclear density and potentially rule-out certain `soft' equations of state \cite{lp07}.  The companion mass is compatible with those of a NS, WD, or main-sequence (MS) companion. Although the spin parameters of \psr\ (Table~1) resemble those of other Galactic disk MSPs, the pulsar is clearly distinct when orbital eccentricity is compared as well (Fig.~2).  Further details of the search pipeline, follow-up observations, and the timing analysis are provided in the supporting online material.

Given the possibility that the companion could be a NS, and potentially a pulsar, we searched for pulsations from the companion using several of the Arecibo observations, but found none. These null results set an upper limit on pulsed emission at 1.4\,GHz of $\sim$20\,$\mu$Jy for a period of 2\,ms and $\sim$9\,$\mu$Jy for a period of 200\,ms assuming a pulse duty cycle of 30\% of the pulse period.

To search for a MS companion, we obtained images of the pulsar field with the Gemini North telescope on July 24, 2007. The total exposure times were 10\,min in the infrared J, H and K$_S$ bands (1.27, 1.67 and 2.22~$\mu$m, respectively). After calibrating the astrometry and photometry of the images against the 2MASS catalog, we find a single star within the 0.13\asec 1-$\sigma$ frame-tie error circle at the position of the pulsar (Fig.~3). It has $J=19.22(9)$, $H=18.41(10)$ and $K_{\rm{S}}=18.03(9)$ magnitudes. Given the density of stars in this field, we estimate the probability of finding a star in the error circle by chance is 2.6\%.  Using MS star models \cite{gbbg00} and estimating the reddening with red clump stars \cite{lcgh02} at the $\sim$6.4\,kpc distance inferred from the pulsar's DM \cite{cl02}, we find that a 0.9\,M$_\odot$ star of age 10\,Gyr would have similar magnitudes to those observed.  The uncertainties in the distance and reddening measurements also allow for a 1.05\,M$_\odot$ star of age 1--5\,Gyr making it possible that this (likely) MS star is a companion to \psr.

If the MS star is the pulsar's binary companion, ionized stellar winds might be detected as increases in the measured DM or as an eclipse of the pulsed flux when the companion passes between the pulsar and Earth. From multi-frequency observations around the orbit, we find no evidence of an eclipse, and we limit any additional DM contribution to $<$0.02\,pc\,cm$^{-3}$.  If the companion of \psr\ is a $\sim$1\,M$_\odot$ main-sequence star with Solar-like winds, we would expect an additional DM contribution of order 10$^{-3}$\,pc\,cm$^{-3}$ near conjunction\cite{yhc+07}.  Strong irradiation of the companion by the pulsar's relativistic wind, however, could lead to a substantially larger mass loss.  Our DM variation limit argues against such a large mass loss.

For a system like \psr, a rotationally induced quadrupole moment in the companion could cause a classical periastron advance \cite{wex98} that would contribute to the measured $\dot\omega = 2.46(2)\times10^{-4}$\,deg\,yr$^{-1}$.  If the companion star is a WD rotating near breakup velocity, the classical contribution would typically be of order 10$^{-7}$\,deg\,yr$^{-1}$, but could be more than an order of magnitude larger for specific but unlikely system orientations.  If the companion is a 1$-$2\,Gyr, 1.05\,M$_\odot$ main-sequence star with a typical rotational period of 8$-$10 days \cite{cc07}, the classical contribution would be 2$\times$10$^{-6}$\,deg\,yr$^{-1}$ for most system orientations.  Such a star would need a rotational period between 1.3 and 1.5 days and/or an unlikely system orientation to account for $\sim$10\% of the measured $\dot\omega$, or a rotational period $<$0.5 day to account for all it. These numbers suggest that the measured $\dot\omega$ is dominated by general relativistic effects and, given the high quality of the fit, that the use of the relativistic timing model is well justified.

\section*{Formation Mechanisms}

What is the origin of this unique system with a short spin period, large orbital eccentricity, and possible MS companion?  According to conventional evolutionary scenarios \cite{Sta04}, binary pulsars that have been recycled down to millisecond periods should always appear in circularized orbits.  In contrast, pulsars in eccentric systems should be only mildly recycled or not recycled at all.  Since \psr\ does not fall into either of these broad categories, we consider three alternative scenarios. The first is that the pulsar was not recycled but was born spinning rapidly in an eccentric orbit at the time the NS was created.  The second has the pulsar recycled in a globular cluster and then ejected into the Galactic disk.  The third has the pulsar recycled in a hierarchical triple system.

For various reasons it seems unlikely that the pulsar was formed spinning rapidly at the time of core-collapse with a small surface magnetic field (2$\times$10$^8$\,G).  First, there are no pulsars like \psrnum\ in any of the more than 50 young supernova remnants in which a NS has been inferred or detected directly \cite{kh02}.  Second, a ``born-fast'' scenario for \psr\ would likely be expected to account for some or all of the 18 isolated MSPs detected in the Galactic disk. While the formation of those systems is puzzling since the observed timescales for evaporating a companion star seem too long, their spin distributions, space velocities, and energetics are indistinguishable from those of recycled (i.e.~not born-fast) binary MSPs and their space velocities and scale heights do not match those of non-recycled pulsars \cite{lmcs07}.  Third, magnetic fields in young pulsars likely originate either from dynamo action in the proto-NS \cite{td93a} or via compression of `frozen-in' fields of the progenitor star during collapse \cite{fw06}.  If compression is the correct mechanism, then young pulsars with magnetic fields $<10^{10}$\,G are rare, as we know of none.  Alternatively, the dynamo model actually requires rapidly spinning systems to have strong magnetic fields. While core-collapse born-fast mechanisms seem to be ruled out, the accretion induced collapse of massive and rapidly rotating WDs to NSs might form MSPs \cite{fw07}.  This collapse may be able to produce the observed orbital parameters, but the large observed pulsar mass would require the collapsing WD to be well above the Chandrasekhar mass, and would also suggest that the companion should be evolved.

Globular clusters (GCs) are known to be efficient producers of MSPs, including those in eccentric binaries, due to interactions between NSs and other stars or binaries in their high density cores.  Of the $\sim$130 known GC pulsars \cite{GCPSRs}, more than 10\% are in highly eccentric ($e>0.2$) orbits.  These numbers, combined with the known populations of NSs in GCs and the Galaxy, and the respective masses of GCs and the Galaxy, imply that GCs produce eccentric binary pulsars at least 1000 times more efficiently per unit mass than the Galactic disk.  Furthermore, stellar interactions and exchanges can provide MS companions for MSPs, although those companions should be less massive than the most evolved MS stars currently observed in GCs ($\sim$0.8$-$0.9\,M$_\odot$).

Although GCs seem a natural formation ground for \psr, there is no known GC located near \psr\ on the sky, nor is there any evidence for an unknown cluster in the 2MASS catalog \cite{scs+06}, the {\em  Spitzer} GLIMPSE survey data \cite{bcb+03}, or in our Gemini observations.  An intriguing possibility is that \psr\ was formed in the core of a GC and then ejected from the cluster, possibly in the same interaction that induced the orbital eccentricity we observe today. This would have to have occurred sometime within the last $\sim$1$-$2\,Gyr (i.e.~within the characteristic age of the MSP). Detailed simulations of NS interactions in GCs have shown that up to $\sim$50\% of recycled NSs are ejected within the $\sim$10\,Gyr lifetimes of the GCs \cite{ihr+07}.  Once the pulsar has left the cluster it drifts far away from its parent cluster on 10$^8$$-$10$^9$ year timescales.  Alternatively, the cluster could have been disrupted during orbital passages through the Galactic disk and bulge within the past $\sim$1\,Gyr \cite{go97}.  Rough estimates based on the masses and densities of the GC system and the Galactic disk suggest (see the supporting online material) a 1$-$10\% chance that \psr\ originated in a GC.  The characteristic age of the pulsar is perhaps the biggest challenge for a GC explanation as it requires that the newly-made MSP had one or more violent interactions which replaced the star that recycled the NS, and then was displaced enough from the GC such that the GC is no longer visible to us, all within $\sim$2\,Gyr.

In a third formation scenario \psr\ is part of a primordial hierarchical triple system. In such a system the inner binary evolves normally. In this case the secondary the secondary mass likely needs to be quite fine-tuned to to produce a highly-recycled massive MSP while still producing a high-mass (0.9$-$1.1\,M$_\odot$) WD in a wide circular orbit, although no such borderline systems have been observed to date. In this scenario the WD is the companion seen in the timing. The third star in the system, now the MS star that we detect as the infrared counterpart, is in a much wider and highly inclined orbit around the inner binary. Secular perturbations in such a system can cause large oscillations of the inner-binary eccentricity and the outer-orbit inclination (so-called Kozai cycles) \cite{koz62}. Especially for a cycle period resonant with the inner-binary relativistic periastron advance at $\sim 2 \times 10^{6}$\,yrs, a 0.9\,M$_{\odot}$ MS star $\sim$120\,AU out in a highly-inclined $\sim700$\,yr orbit can induce large inner-binary eccentricities\cite{fkr00}. Initial estimates (see the supporting online material) for the formation and survival probability of such a triple system suggest that a few percent of observed NS--WD binaries could be members of hierarchical triples.  As such, it seems plausible that after finding $\sim$50 pulsar--WD binaries, we have now found the first in an hierarchical stellar triple.

A related scenario that avoids the need for formation of both a fully recycled MSP and a high-mass WD companion (which has never been observed before) has \psr\ recycled as part of a compact inner binary in a hierarchical triple, in a configuration as recently suggested for 4U~2129$+$47 \cite{btgc08}.  The MSP then ablated away its WD companion and was left in a 95-day eccentric orbit around the MS star we now observe.

Further observations of \psr\ will allow us to decide between these (or other) formation scenarios.  A measurement of a large projected space velocity ($>$100-200\,km\,s$^{-1}$), via long-term timing or Very Long Baseline Interferometry astrometry, might reflect a cluster origin given the high velocities of most GCs.  Spectroscopic observations of the MS star will reveal its spectral type and metallicity, both possible indicators for or against a GC origin, and will show whether it exhibits the 95-day orbital motion of the pulsar. If the star is the companion, the radial-velocity curve will further constrain the masses of both the pulsar and the companion. Additionally, if the MS companion is confirmed to be more massive than $\sim$1\,M$_\odot$, it will likely rule out a GC origin as MS stars of that mass in clusters have already left the MS.  Finally, long-term and higher-precision timing of the pulsar will dramatically improve the relativistic parameters of the system (and therefore the derived masses) and will reveal secular changes in the spin and orbital parameters caused by the presence of a third star or by classical effects from the rotation of a MS companion.

\bibliography{1903}
\bibliographystyle{Science}

\begin{scilastnote}
\item We thank the staff at NAIC and ATNF for developing ALFA and its associated data acquisition systems. This work was supported by the NSF through a cooperative agreement with Cornell University to operate the Arecibo Observatory. The National Radio Astronomy Observatory is a facility of the National Science Foundation operated under cooperative agreement by Associated Universities, Inc. Pulsar research at Cornell is supported by NSF grants AST 0507747 and CISE/RI 040330 and by the Center for Advanced Computing. The McGill pulsar group Beowulf computer cluster used for this work was funded by the Canada Foundation for Innovation. Pulsar research at Bryn Mawr College is funded by NSF grant AST 0647820. Basic research in radio astronomy at NRL is supported by 6.1 Base funding. Pulsar research at Columbia University is supported by NSF grant AST 0507376. Pulsar research at UBC is supported by an NSERC Discovery Grant.  V.\,M.\,K. holds a Canada Research Chair and the Lorne Trottier Chair and acknowledges support from an NSERC Discovery Grant, CIFAR, and FQRNT. J.\,W.\,T.\,H.~holds an NSERC Postdoctoral Fellowship and CSA supplement. L.\,K. holds an NSERC CGS-D fellowship.  J.\,v.\,L.~is a Niels Stensen Fellow. Pulsar research at the University of Texas at Brownsville is funded by NSF grant AST 0545837. We thank O. Pols, D. Fabrycky and G. Duch\^ene for helpful discussions.
\end{scilastnote}

\clearpage

\begin{table*}
\begin{center}
\begin{tabular}{l l}
  \hline
  \multicolumn{2}{c}{Timing Parameters Assuming General Relativity} \\
  \hline
  Right Ascension (J2000) \dotfill 					& 19$^{\rm h}\;$03$^{\rm m}\;$05$\fs$79368(3) \\
  Declination (J2000) \dotfill 						& 03$\degrees\;$27$\amin\;$19$\farcs$2220(11) \\
  Spin Period (ms) \dotfill 							& 2.1499123298435(3) \\
  Period Derivative (s/s) \dotfill 						& 1.879(2)$\times$10$^{-20}$ \\
  Dispersion Measure (pc cm$^{-3}$) \dotfill 			& 297.537(7) \\
  Epoch of Period (MJD) \dotfill 						& 54280.0 \\
  Orbital Period (days) \dotfill 						& 95.1741176(2) \\
  Projected Semi-Major Axis (lt-s) \dotfill 				& 105.60585(11) \\
  Eccentricity \dotfill 								& 0.436678411(12) \\
  Longitude of Periastron \dotfill 					& 141.65779(4)$\degrees$ \\
  Epoch of Periastron (MJD) \dotfill 					& 54063.8402308(5) \\
  Total System Mass, M$_{\rm tot}$ (M$_\odot$) \dotfill 	& 2.79(5)\\
  Companion Mass, M$_2$ (M$_\odot$) \dotfill 		& 1.051(15)\\
  \hline
  \multicolumn{2}{c}{Other Parameters} \\
  \hline
  Scattering Time at 1.4\,GHz (ms) \dotfill 				& 0.126(1)\\
  1.4-GHz Flux Density (mJy) \dotfill 					& 1.3(4)\\
  2-GHz Flux Density (mJy) \dotfill 					& 0.62(5)\\
  5-GHz Flux Density (mJy) \dotfill 					& 0.09(2)\\
  Spectral Index \dotfill 							& -2.1(2)\\
  \hline
  \multicolumn{2}{c}{Derived Parameters} \\
  \hline
  Galactic Longitude (J2000) \dotfill 					& 37\fdg3363\\
  Galactic Latitude (J2000) \dotfill 					& $-$1\fdg0136\\
  Mass function (M$_\odot$) \dotfill 					& 0.1396076(2)\\
  Distance (kpc) \dotfill							& $\sim$6.4\\
  Surface Dipole Magnetic Field Strength (Gauss) \dotfill 	& 2.0$\times$10$^8$\\
  Characteristic Age (Gyr) \dotfill 					& 1.8\\
  Spin-down Luminosity (ergs\,s$^{-1}$) \dotfill 			& 7.5$\times$10$^{34}$\\
  Advance of Periastron (deg yr$^{-1}$) \dotfill 			& 2.46(2)$\times$10$^{-4}$\\
  Orbital inclination \dotfill 							& 78(2)$\degrees$ \\
  Pulsar Mass, M$_1$ (M$_\odot$) \dotfill 				& 1.74(4)\\
  \hline
\end{tabular}

\caption{\label{tab:ephemeris}Measured and derived parameters for \psr.  The timing parameters were measured using the DE405 Solar System ephemerides \cite{DE405} and the ``DDGR'' timing model which assumes that general relativity fully describes the parameters of the binary system \cite{tw89}.  A total of 342 pulse arrival times measured between MJDs 53990 and 54568 were fit.  The numbers in parentheses are twice the TEMPO-reported 1-$\sigma$ uncertainties in the least significant digit or digits quoted as determined by a bootstrap error analysis.  The distance is inferred from the NE2001 free electron density model \cite{cl02}. Ensemble distance measurements using this model have an estimated error of 25\%, although errors for individual pulsars may be larger.}
\end{center}
\end{table*}

\clearpage

\begin{figure}[hb]
\begin{center}
\includegraphics[height=5in,angle=0]{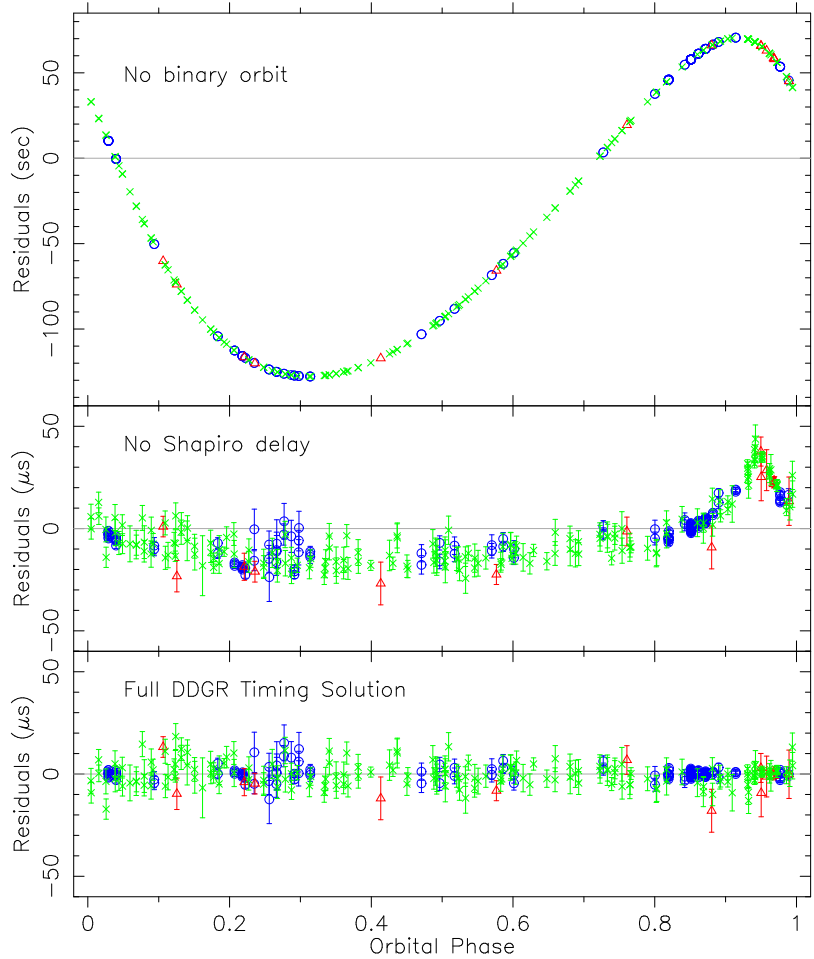}
\end{center}
\caption{Residual pulse arrival times as a function of orbital phase (mean anomaly) for \psr\ after subtraction of the best-fit timing model.  The timing residuals are from observations made with the Arecibo telescope (blue circles), the Westerbork Synthesis Radio Telescope (red triangles), and the Green Bank Telescope (green crosses), and are defined as observed minus model. {\em Top} The measured timing residuals if no orbit is accounted for.  The resulting curve is the Roemer delay (i.e.~the light-travel time across the orbit) and its non-sinusoidal shape shows the large eccentricity ($e = 0.44$) of \psr.  The uncertainties on the data points would be the same as in the lower panels but the scale is different by a factor of 10$^6$.  {\em Middle} The same residuals as in the top panel but with the Roemer delay and all general relativistic delays except for Shapiro delay from the timing solution in Table 1 removed.  {\em Bottom} The timing residuals for the full ``DDGR'' timing model described in Table~1 which assumes that general relativity fully describes the parameters of the binary system \cite{tw89}. The weighted root-mean-square timing residual shown here is 1.9~$\mu$s. }
\end{figure}

\begin{figure}[hb]
\begin{center}
\includegraphics[height=5in,angle=0]{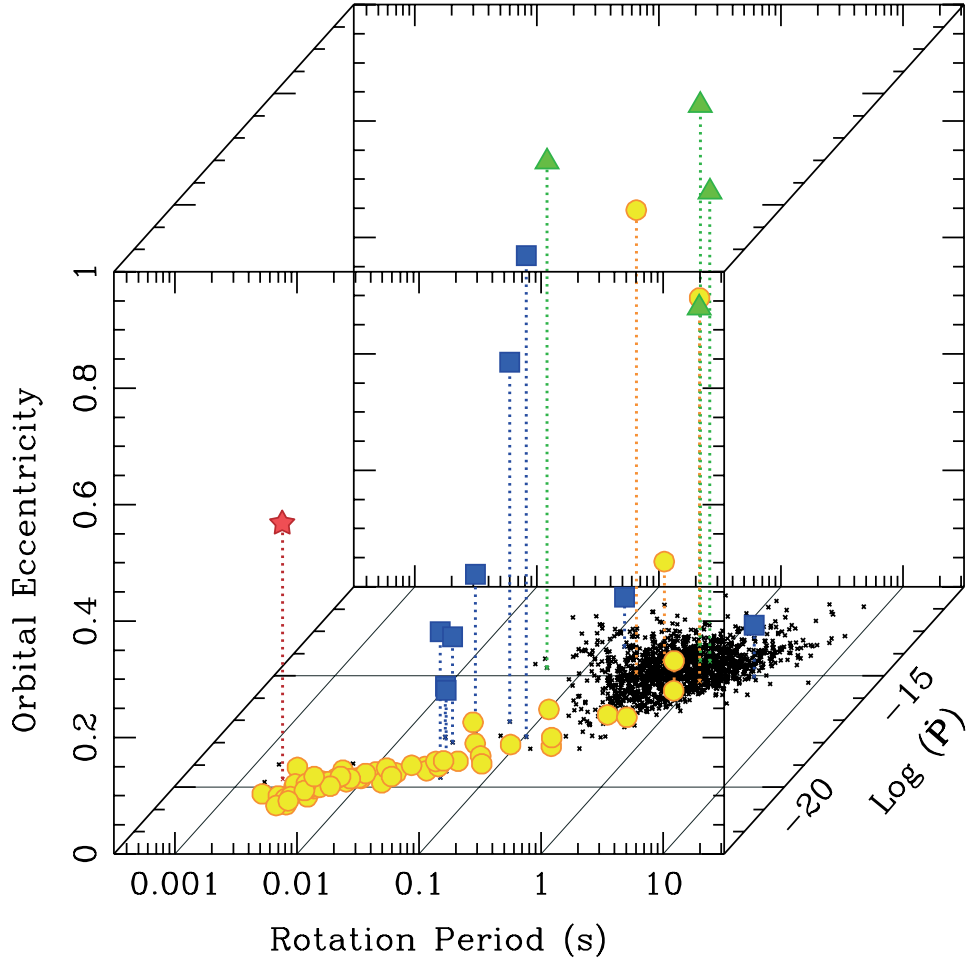}
\end{center}
\caption{Rotation periods, period derivatives, and orbital eccentricities (for binary pulsars) of pulsars in the disk of the Galaxy.  The bottom face of the cube shows a plot of rotation period versus rotation period derivative for all Galactic pulsars.  Colored points show the binary pulsars, projected upward from the bottom face in proportion to their orbital eccentricities.  Square blue points are double neutron star systems, triangular green points are pulsars with main-sequence or massive companions, circular yellow points are pulsars with white dwarf or sub-dwarf companions, and the red star is PSR~J1903+0327, which occupies a unique place in the diagram.}
\end{figure}

\begin{figure}[hb]
\begin{center}
\includegraphics[height=5in,angle=0]{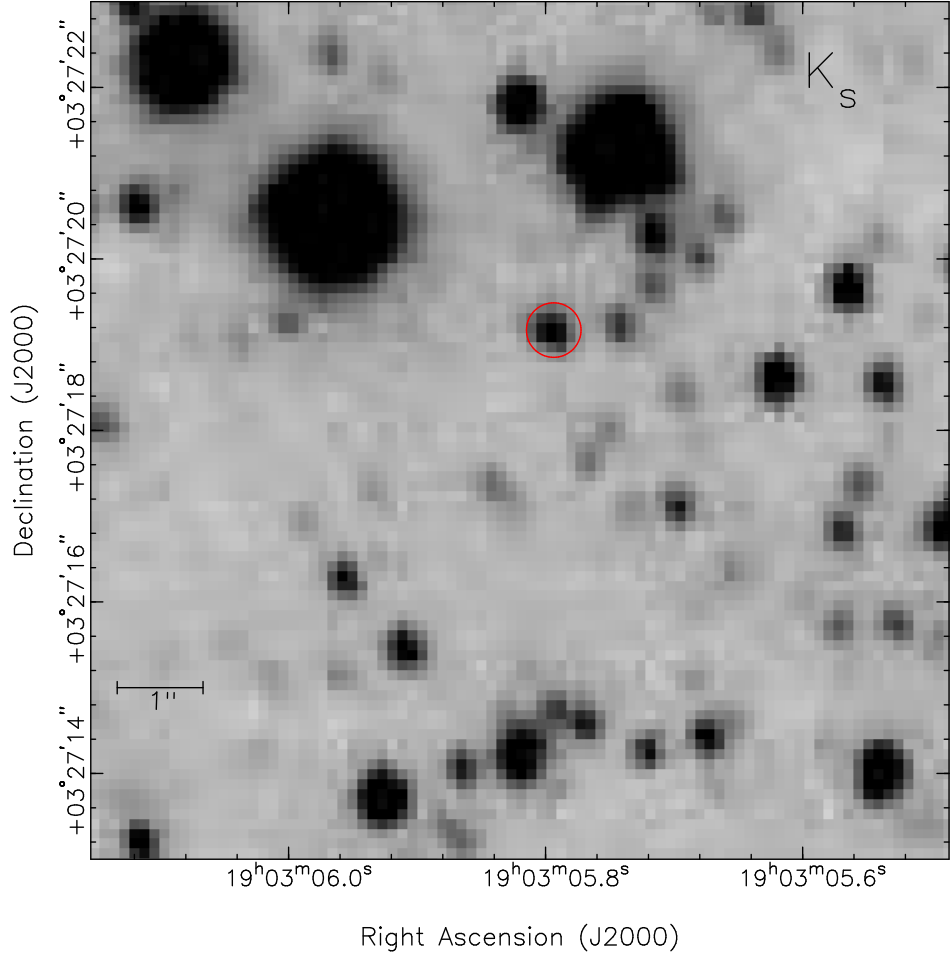}
\end{center}
\caption{A K$_S$-band image of the \psr\ field  taken during excellent seeing conditions (0.3$-$0.4\asec) with the Gemini North telescope. The red circle shows the 2-$\sigma$ error circle, with radius 0\farcs32 (produced by the frame-tie uncertaities in right ascension and declination), for the position of the pulsar based on astrometric calibrations made with the 2MASS catalog. The star within the error circle is the possible main-sequence companion to the pulsar.}
\end{figure}

\renewcommand\citeform[1]{S#1}
\makeatletter
	\renewcommand{\@biblabel}[1]{S#1.}
	\renewcommand{\thetable}{S\arabic{table}}
	\renewcommand{\thefigure}{S\arabic{figure}}  
\makeatother

\setcounter{table}{0}
\setcounter{figure}{0}

\section*{Materials and methods}

In the following, we describe several details of the discovery and follow-up observations of \psr\ as well as statistical arguments relating to origin in a globular cluster or a triple star system.

\subsection*{Discovery}

The survey \cite{cfl+06} collects data during 5-min pointings for each of seven positions on the sky using the Arecibo L-band Feed Array (ALFA) receiver and digital spectrometers covering 100-MHz of bandwidth centered at 1400\,MHz. The recorded data have 256 frequency channels and a time resolution of 64\,$\mu$s. We initially detected \psr\ in September 2006 using a search pipeline based on the PRESTO suite of pulsar analysis software\cite{rem02,presto}.  The basic steps of the pipeline are similar to those of the ``quicklook pipeline'' \cite{cfl+06}, except that we process the data at full resolution in both time and radio frequency, we perform extensive radio-frequency interference excision in the time and frequency domains, and we look specifically for compact binary pulsars using linear ``acceleration'' searches \cite{lk05}. To remove the dispersive effects of the interstellar medium, the data were first dedispersed for 1272 trial dispersion measures (DMs) between 0$-$1003\,pc\,cm$^{-3}$.  The DM trial values are spaced such that the pulse smearing due to interstellar dispersion is $<$1\,ms for all DMs $<$600\,pc\,cm$^{-3}$, thereby giving us unprecedented sensitivity to millisecond pulsars (MSPs) in the disk of our Galaxy.  We conduct standard periodicity and acceleration searches by taking the Fast Fourier Transform (FFT) of each dedispersed time series using incoherent harmonic summing of up to 16 harmonics. In addition, single-pulse searches of the data were carried out which were sensitive to transient events in the time domain with widths in the range 64\,$\mu$s to 0.1\,s

\subsection*{Timing}

Once a pulsar is discovered it is characterized by its position, spin parameters and any orbital parameters. Using the search data these parameters are poorly constrained, but they can be measured to much greater levels of precision by fitting a timing model of the pulsar's rotation to the data. Integer numbers of pulsar rotations are fitted between the times-of-arrival of the observed pulses; in such a ``phase connected solution'' every rotation of the pulsar is accounted for.

Follow-up timing observations were carried out with Arecibo at $\sim$1.5\,GHz on 22 days between September 2006 and January 2007 and at $\sim$2.1\,GHz for several days between January 2008 and April 2008, with the Westerbork Synthesis Radio Telescope (WSRT) at $\sim$1.4\,GHz on 14 days between April and December 2007, and with the Green Bank Telescope (GBT) primarily at $\sim$2.1\,GHz between December 2006 and April 2008. The Arecibo observations used the seven beam ALFA receiver and a Wideband Arecibo Pulsar Processor (WAPP) usually in the same configuration as used in the survey.  The GBT observations made use of the Green Bank Pulsar Spigot \cite{kel+05}, a digital correlator which synthesized either 768 or 1536 frequency channels covering 600\,MHz of bandwidth and sampled every 81.92\,$\mu$s. Individual GBT observations were made at 5 and 9\,GHz with the Spigot in a mode with 1024 frequency channels recorded over 800\,MHz of bandwidth and sampled every 81.92\,$\mu$s.  WSRT observations used the PUMa II data acquisition system \cite{kss08} to record eight 20\,MHz wide dual polarization bands.  Each band was subsequently analyzed offline by forming a 64-channel coherent filterbank leading to a final time resolution of 16.8\,$\mu$s.

The timing model parameters shown in Table 1 of the main article were derived from the best-fit model to the pulse arrival times we obtained with the above telescopes and instrumentation.  A timing model using a simple Keplerian orbit provides a very poor fit to the data (reduced-$\chi^2$ of 29.5 for 327 degrees of freedom [DOF]).  The addition of an orbital periastron advance ($\dot\omega$) dramatically improves the fit (reduced-$\chi^2$ of 2.12 for 326 DOF), giving $\dot\omega = 2.46(2)$$\times$10$^{-4}$\,deg\,yr$^{-1}$, but systematic trends remain in the timing residuals. Given the orientation of the orbit, these remaining trends are most likely explained by the Shapiro delay: the delay should have an amplitude of tens of $\mu$s, an order of magnitude larger than our arrival time measurement precision, even though most of that signal will be covariant with the projected semi-major axis, $a\sin i$. Using the ``DDGR'' model \cite{tw89,dd86}, which assumes that general relativity correctly describes the dynamics of the system with Keplerian orbit parameters plus the total system mass M$_{\rm tot}$ and the mass of the companion star M$_2$, the fit is further improved (reduced-$\chi^2$ = 1.13 for 325 DOF) and results in no visible systematics in the residuals (Fig.~1).  In Fig.~S1, we show a $\chi^2$ contour map of the mass constraints from the DDGR model.

In order to confirm that this timing is consistent with general relativity, the ``DD'' model \cite{dd85,dd86}, a theory independent relativistic description of the orbital parameters, was used separately fitting for the orbital periastron advance, the companion mass and sine of the inclination angle ($\sin$i). In this fit (reduced-$\chi^2$ = 1.12 for 324 DOF) all of the post-Keplerian parameters which are not specific to general relativity are consistent with the parameters from the general relativity specific DDGR model at the 1-2\,$\sigma$ level. The timing model parameters for the DD model are show in Table~S1.

\subsection*{Search for a pulsar companion}

Given the possibility that this is a double neutron star system, and that the companion is potentially a pulsar, we searched for pulsations from the companion using several of the early observations of the system taken at Arecibo. As well as searching each observation independently, we summed power spectra from the individual observations to increase sensitivity.  We dedispersed each observation at the known DM of \psr.  We then removed the deleterious orbital effects of the system prior to incoherently summing the power spectra from FFTs. The orbital accelerations were calculated and removed by noting that the orbital period and eccentricity of the companion's orbit are the same as for \psr, but the longitude of periastron is different by 180\degrees.  The only unknown parameter in this calculation is the ratio of the semi-major axes of the pulsar and companion which is the inverse of the ratio of their masses. A search over a series of trial projected semi-major axes for the companion was therefore used, ranging from 60 to 185 light-seconds. This encompasses the range of well-measured neutron star masses \cite{lp07}.  Finally the summed FFT was searched for candidates using the techniques described above. The search yielded no significant candidates.

We set an upper limit on pulsed emission at 1.4\,GHz of 20\,$\mu$Jy for a period of 2\,ms and 9\,$\mu$Jy for a period of 200\,ms, both at a DM of 297\,pc\,cm$^{-3}$.  These upper limits correspond to 1.4\,GHz radio luminosities of 0.8 and 0.4~mJy~kpc$^2$ respectively. Only two percent of all radio pulsars currently known have luminosities below the latter value.  Our non-detection therefore excludes most of the observable radio pulsar population as a companion, but does not of course rule out a neutron star that is unfavorably beamed away from our line of sight or rotating too slowly to be observed as a pulsar.

\subsection*{Optical and infra-red analysis}

The field of PSR~J1903+0327 was observed with the Near InfraRed Imager (NIRI) on the Gemini North telescope on July 24, 2007. Dithered series of 6\,second exposures were obtained in the $J$, $H$ and $K_\mathrm{S}$ filters, amounting to a total exposure time of 10\,minutes in each filter.  All images were corrected for dark current and flatfielded using skyflats. The non-uniform sky distribution was removed by substracting a sky frame constructed of the unregistered science images to remove the contributions of stars. After these corrections, the images taken with the same filter were registered and averaged.

Astrometry was done relative to the 2MASS catalogue \cite{scs+06}, as the small field of view of NIRI ($2\amin\times2\amin$) contains no astrometric standards from the UCAC2 catalogue \cite{zuz+04}. A total of 71 2MASS stars coincided with the registered and average $K_\mathrm{S}$-band image and 51 of these were not saturated and appeared steller and unblended. After iteratively removing 9 outliers, the final astrometric calibration has rms residuals of $0\farcs086$ in right ascension and $0\farcs094$ in declination.  A comparison of the positions of 281 UCAC2 standards coinciding with 2MASS stars within $10\amin$ from the position of PSR~J1903+0327 shows no significant shift between the positions in both catalogues.

Taking the rms uncertainty on both coordinates as a measure for the 1-$\sigma$ uncertainty in the astrometric calibration, we find a single star inside the $0\farcs13$ error circle on the pulsar position at $\alpha_{\rm{J2000}}=19^{\rm{h}}03^{\rm{m}}05\fs796(6)$ and $\delta_{\rm{J2000}}= 03\degrees27\amin19\farcs19(9)$, where the uncertainty on the position is the quadratic sum of the uncertainty in the astrometry and the intrinsic positional uncertainty of the star in the image ($0\farcs02$ in each coordinate). This position if offset from the pulsar position by $0\farcs034(88)$ in right ascension and $0\farcs032(96)$ in declination.

Photometry was also done relative to the 2MASS catalogue. Magnitude offsets were determined and outliers were iteratively removed, leaving 30 to 40 stars for the photometric calibration. The calibrations showed no significant dependence with star color. We find that the object in the error circle has $J=19.22(9)$, $H=18.41(10)$ and $K_{\rm{S}}=18.03(9)$.

To estimate the reddening towards PSR~J1903+0327, we used red clump stars to trace the reddening as a function of distance \cite{lcgh02}. Again using the 2MASS catalogue, we selected 1400 stars within $5\amin$ from PSR~J1903+0327 and determined the $J-K_\mathrm{S}$ colour of the red clump stars at different $K_\mathrm{S}$ magnitudes using the formalism of \cite{dv06}. At the DM-distance of PSR~J1903+0327, $d=6.4(16)$\,kpc (assuming a 25\% uncertainty), we constrain the reddening at $A_V=4.9(11)$. For this distance and reddening, the intrinsic magnitudes of the star inside the error circle are $M_J=3.8(11)$, $M_H=3.5(10)$ and $M_{K_{\mathrm{S}}}=3.5(10)$. 

Though the metallicity and age of the star are unknown, stars of solar metallicity, ages of 1\,Gyr or less and masses between approximately 0.8 and 1.3\,M$_\odot$ will have similar absolute magnitudes \cite{gbbg00}. Stars older than 1\,Gyr will have evolved and will match the absolute magnitudes for lower masses. At 10\,Gyr and solar metallicity the mass range decreases to approximately 0.75 to 1.1\,M$_\odot$. For a mass of 1.05\,M$_\odot$ as determined from the radio timing, a solar metallicity star will match the observed absolute magnitudes if the age is less than 6\,Gyr.

At the apparent magnitudes, the star is bright enough for optical or infrared spectroscopy to determine radial velocity variations which would unambiguously confirm the star as the binary companion to PSR~J1903+0327. 

\subsection*{Globular cluster calculations}

While \psr\ is unprecedented in the Galactic plane several MSPs in eccentric binaries have been seen in globular clusters (GCs). GCs are known to be excellent breeding grounds for MSPs, with formation rates per unit mass $\sim$100 times that in the disk \cite{fpr75,Ver87}. \psr\ may have formed in a GC which was later disrupted, or escaped from an existing GC $\sim 10^9$~yr ago such that they are no longer in the same part of the sky. Many GCs may have been completely disrupted via gravitational interactions with the disk and bulge over the course of many orbits in the Galactic potential. As a result, their stars become part of the Galactic spheroid population (an approximately spherical distribution of older stars distributed in an extension of the central Galactic bulge, with a diameter of $\sim$10\,kpc). Estimates suggest that over half the spheroid mass could have come from such disrupted GCs \cite{go97}.  Such disruption should require several Galactic orbits of the GC, the periods of which are $\sim$0.5$-$1.0 Gyr.  However, in the particular case of \psr, we are constrained by the relatively small characteristic age of the system. Since a characteristic age is usually an upper limit on the true age of a radio pulsar \cite{lk05}, any model involving GC disruption requires a recycled pulsar to exchange its companion and acquire an eccentric orbit and its parent GC to then be totally disrupted in less than $\sim$2\,Gyr. This is challenging given the orbital periods of GCs.

By comparing the densities of the spheroid and disk we can estimate the probability of \psr\ being found in the disk but actually being associated with the spheroid population. Using estimates of the stellar densities at low Galactic latitudes of both the spheroid \cite{rrc00} and disk \cite{Chab01} populations, we find $\rho_{\rm  spheroid}/\rho_{\rm disk}=4.7\times10^{-5}{\rm\ M_\odot\  pc^{-3}}/4.3\times10^{-2}{\rm\ M_\odot\ pc^{-3}} \sim 1\times10^{-3}$.  Assuming that half of the spheroid mass originated in disrupted GCs, the $\la$2\,Gyr age of \psr\ implies that $\la$20\% of that mass could have come from GCs disrupted since the recycling of the pulsar. Furthermore, if we assume that those GCs produced highly eccentric binary MSPs $\sim$10$^3$ times more efficiently per unit mass than the disk, we can crudely estimate that the probability that \psr\ came from a disrupted GC is $\la$10\%.

We can make a similar estimate of the probability that \psr\ was ejected from a GC.  The number of GC-ejected \psr-like systems should be roughly the ratio of the GC system mass (10$^7$\,M$_{\odot}$) to the disk mass (10$^{11}$\,M$_{\odot}$) times the 10$^3$ formation efficiency factor for highly eccentric binary MSPs times the $\sim$50\% fraction of pulsars which are eventually ejected from GCs \cite{ihr+07}. The resulting $\sim$5\% probability again indicates that some GC formation mechanism for \psr\ is plausible.

\subsection*{Hierarchical triple system calculations}

Hierarchical stellar triple systems are common, and such a system may explain the eccentricity we observe in \psr. To date no pulsars have been found in other stellar triples. Hierarchical triples contain an inner binary (in this case the MSP and the $\sim$1.05\,M$_{\odot}$ companion seen in timing at the 95-day orbit) with a third object orbiting at greater distance (in this case, the main sequence star observed in the infra-red). Here we assess the plausibility of forming such an end state.

Constraints on the formation of the inner binary are: (i) to form a 2.15-ms MSP there must have been a long period of stable mass transfer to spin up the neutron star; (ii) to produce the wide 95-day orbit the white dwarf progenitor cannot be too massive, to avoid common-envelope (CE) evolution and spiral-in; (iii) the star must have been massive enough to leave a $\sim$1.05\,M$_{\odot}$ white dwarf after final mass transfer \cite{tvs00}. We therefore consider an initial system containing a central binary formed by a 8\,M$_{\odot}$ primary; a secondary that is just below CE mass, we shall here assume 3\,M$_{\odot}$; and a 0.9\,M$_{\odot}$ third star.

A survey of the multiplicity of massive stars in the Orion Nebula \cite{pbh+99} suggests that 80\% of spectroscopic binaries with OB-star primaries have a third visual companion. These systems must survive the supernova that creates the pulsar: the inner binary will remain bound given suitable mass-loss and kick parameters \cite{Hill83}. The outer companion can remain bound \cite{py06} if its orbital velocity is comparable to the new system velocity of the inner binary (a few tens of km\,s$^{-1}$, \cite{mhth05}). The multiplicity survey \cite{pbh+99} finds $\sim$5-10\% of third companions are close-in enough to remain bound in the supernova. The semi-major axis of this outer orbit may then increase as mass is lost from the system during the pulsar spin-up phase \cite{spv97}. In this hierarchical triple the outer companion cannot unbind the inner binary, but its orbit evolves towards a Kozai-cycle resonance with the inner-binary relativistic periastron advance period\cite{fkr00}. Our above rough estimates of the various fractions combine to a roughly 4\% chance that a given evolved NS-WD binary is part of a hierarchical triple. As there are currently $\sim$50 NS-WD systems known in Galactic-disk binaries, this simple estimate is consistent with the idea that \psr\ is the first in such a triple.  We do note that this formation scenario still faces the difficulty that thus far no other such systems with both a highly recycled pulsar and a wide-orbit very massive WD companion have been observed.

Two other probability arguments for the detection of this system are not directly related to the formation plausibility. First, for a system with masses 1.74, 1.05 and 0.9\,M$_{\odot}$, a relatively high inclination of the third star's orbit $i > 45$\degrees and some eccentricity in the outer orbit, the inner-binary eccentricity is higher than $e=0.44$ for about 20\% of the oscillation time \cite{fkr00}. Second, for the line-of-sight acceleration from the third star to be less than our measured $\dot{P}$ this third star must currently be close to plane of the sky, the a priori probability of which is $\sim$5\%.

\bibliography{1903}
\bibliographystyle{Science}

\clearpage

\begin{table*}
\begin{center}
\begin{tabular}{l l}
  \hline
  \multicolumn{2}{c}{Timing Parameters Using the DD Model} \\
  \hline
  Right Ascension (J2000) \dotfill			& 19$^{\rm h}\;$03$^{\rm m}\;$05$\fs$79368(3) \\
  Declination (J2000) \dotfill				& 03$\degrees\;$27$\amin\;$19$\farcs$2220(11) \\
  Spin Period (ms) \dotfill					& 2.1499123298435(3) \\
  Period Derivative (s/s) \dotfill				& 1.879(2)$\times$10$^{-20}$ \\
  Dispersion Measure (pc cm$^{-3}$) \dotfill	& 297.537(7) \\
  Epoch of Period (MJD) \dotfill				& 54280.0 \\
  Orbital Period (days) \dotfill				& 95.1741176(2) \\
  Projected Semi-Major Axis (lt-s) \dotfill		& 105.593459(8) \\
  Eccentricity \dotfill						& 0.43667838(5) \\
  Longitude of Periastron \dotfill				& 141.652477(3)$\degrees$ \\
  Epoch of Periastron (MJD) \dotfill			& 54063.8402310(6) \\
  Advance of Periastron (deg yr$^{-1}$) \dotfill	& 2.46(3)$\times$10$^{-4}$\\
  Sine of Orbital Inclination  \dotfill			& 0.966(10)\\
  Companion Mass, M$_2$ (M$_\odot$) \dotfill	& 1.3(2)\\
  \hline
\end{tabular}

\caption{Measured and derived parameters for  \psr. The timing parameters were measured using the DE405 Solar System ephemerides \cite{DE405} and the ``DD'' timing model which uses a theory independent relativisitc model to describe the parameters of the binary system \cite{dd85,dd86}.  A total of 342 pulse arrival times measured between MJDs 53990 and 54568 were fit.  The numbers in parentheses are twice the TEMPO-reported 1-$\sigma$ uncertainties in the least significant digit or digits quoted as determined by a bootstrap error analysis.}

\end{center}
\end{table*}

\begin{figure}[hb]
\begin{center}
\includegraphics[height=5in,angle=270]{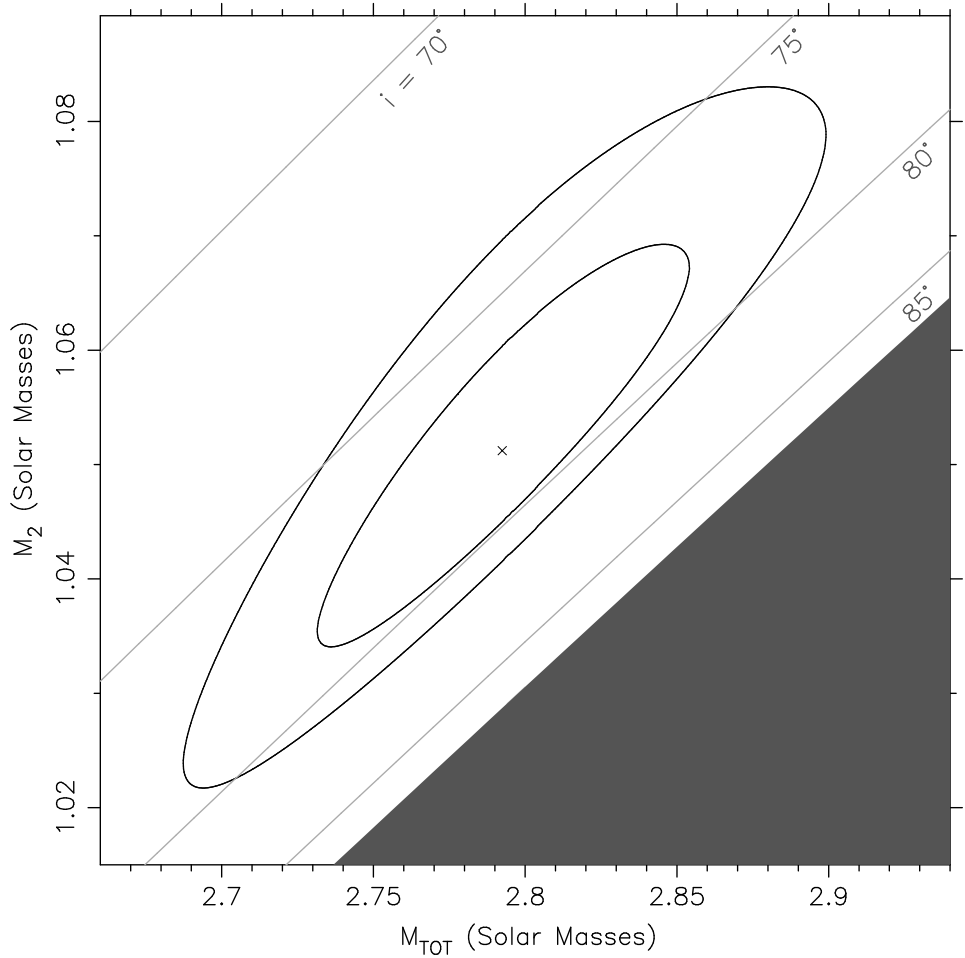}
\end{center}
\caption{A $\chi^2$-map showing the 1 and 2-$\sigma$ confidence regions for the total system mass (M$_{\rm tot}$) and companion mass (M$_2$) based on the timing solution presented in Table~1 after doubling the nominal TEMPO errors. Those errors were within 20\% of the bootstrap error estimates. The grey region in the lower right portion of the figure is excluded by the mass function.  The grey lines show inclinations $i$ of 70, 75, 80, and 85$\degrees$.  For \psr, M$_{\rm tot}$ is best constrained from the measurement of the relativistic advance of periastron $\dot\omega$ while M$_2$ is constrained by a measurement of $\sin i$ via the Shapiro delay (see Fig.~1).}
\end{figure}

\end{document}